\documentclass[
reprint,
superscriptaddress,
 amsmath,amssymb,
 aps,
]{revtex4-2}
\usepackage{appendix}
\usepackage{graphicx}
\usepackage{dcolumn}
\usepackage{bm}

\usepackage[colorlinks=true,linkcolor=blue,anchorcolor=blue,citecolor=blue, filecolor=blue,urlcolor=blue,bookmarksnumbered=true,pdfview=FitB]{hyperref}

\usepackage{subfig}
\usepackage{mathrsfs}
\usepackage{float}
\usepackage{url}
\graphicspath{{figures}}

\allowdisplaybreaks

\begin{document}


\title{Color Lorentz force on a transversely polarized quark in the nucleon}

\author{Ziqi Zhang}
\email{zhangziqi@impcas.ac.cn}
\affiliation{State Key Laboratory of Heavy Ion Science and Technology, Institute of Modern Physics, Chinese Academy of Sciences, Lanzhou 730000, China}
\affiliation{School of Nuclear Science and Technology, University of Chinese Academy of Sciences, Beijing 100049, China}

\author{Chandan Mondal}
\email{mondal@impcas.ac.cn}
\affiliation{State Key Laboratory of Heavy Ion Science and Technology, Institute of Modern Physics, Chinese Academy of Sciences, Lanzhou 730000, China}
\affiliation{School of Nuclear Science and Technology, University of Chinese Academy of Sciences, Beijing 100049, China}

\author{Siqi Xu}
\email{xsq234@impcas.ac.cn}
\affiliation{State Key Laboratory of Heavy Ion Science and Technology, Institute of Modern Physics, Chinese Academy of Sciences, Lanzhou 730000, China}
\affiliation{School of Nuclear Science and Technology, University of Chinese Academy of Sciences, Beijing 100049, China}
\affiliation{Department of Physics and Astronomy, Iowa State University, Ames, IA 50011, U.S.A.}

\author{Xingbo Zhao}
\email{xbzhao@impcas.ac.cn}
\affiliation{State Key Laboratory of Heavy Ion Science and Technology, Institute of Modern Physics, Chinese Academy of Sciences, Lanzhou 730000, China}
\affiliation{School of Nuclear Science and Technology, University of Chinese Academy of Sciences, Beijing 100049, China}

\author{James P. Vary}
\email{jvary@iastate.edu}
\affiliation{Department of Physics and Astronomy, Iowa State University, Ames, IA 50011, U.S.A.}

\collaboration{BLFQ Collaboration}

\date{\today}

 \begin{abstract}
The $x^2$ moment of the genuine twist-3 component of the unpolarized scalar parton distribution function $e(x)$ is related to the average transverse color Lorentz force experienced by a transversely polarized quark in an unpolarized nucleon. This higher-twist quantity probes quark--gluon correlations and provides unique insight into the nonperturbative dynamics of quantum chromodynamics. For the first time, we study the corresponding off-forward matrix element and derive its Lorentz-covariant decomposition in terms of generalized form factors. We calculate these form factors from light-front wave functions obtained by solving the light-front Hamiltonian eigenvalue problem within the Basis Light-Front Quantization framework. Their Fourier transforms provide a spatial picture of the transverse color forces in impact-parameter space. In particular, $\mathscr{F}_2$, whose forward limit is directly related to the twist-3 matrix element $e_2$, can be interpreted as the impact-parameter-space representation of the Boer--Mulders force, providing an intuitive spatial picture of the final-state interaction mechanism underlying the Boer--Mulders effect.
\end{abstract}

\maketitle


{\it Introduction}.---Understanding the spatial structure of hadrons is one of the central goals of modern hadron physics. Over the past decades, form factors (FFs) and generalized parton distributions (GPDs) have provided increasingly detailed information on the internal distributions of charge, momentum, spin, and mechanical properties inside the nucleon \cite{Pacetti:2014jai,Burkert:2023wzr,Diehl:2023nmm}. In particular, the Fourier transform of gravitational form factors (GFFs) has enabled the visualization of pressure and shear-force distributions associated with the quantum chromodynamics (QCD) energy-momentum tensor (EMT), leading to the emerging field of hadron tomography~\cite{Burkert:2023wzr}. These developments have significantly advanced our understanding of the nonperturbative dynamics governing the nucleon structure.

Beyond the mechanical properties encoded in the EMT, quark–gluon correlations at higher twist provide access to dynamical color  forces acting on partons inside hadrons. Among them, the twist-3 quark–gluon operator $\bar q \sigma^{+i} G^{+i} q$ plays a special role. Its forward matrix element is related to the twist-3 moment $e_2$, which has been interpreted as the average transverse color Lorentz force experienced by a transversely polarized quark in an unpolarized nucleon and is closely connected to the dynamical origin of the Boer--Mulders effect~\cite{Burkardt:2008ps,Abdallah:2016xfk,Burkardt:2007xm}. Previous studies that have mainly focused on the color Lorentz force acting on an unpolarized parton inside a transversely polarized proton \cite{Burkardt:2008ps,Aslan:2019jis,crawford2025transverse,liu2025color,Zhang:2026cig}, have not revealed much detail about the spatial distribution of such transverse color forces.

To address this question, it is necessary to go beyond the forward limit and investigate the off-forward matrix element of the twist-3 quark–gluon operator. Such a description naturally introduces generalized FFs that encode the momentum-transfer dependence of the underlying quark–gluon correlations. Their Fourier transforms provide access to impact parameter space (IPS) distributions \cite{burkardt2002impact}, allowing a spatial imaging of the transverse color forces in close analogy with nucleon tomography based on GPDs and electromagnetic/EMT FFs.

Here, we investigate the off-forward matrix element of $\bar q \sigma^{+i} G^{+i} q$ within the Basis Light-Front Quantization (BLFQ) framework~\cite{Vary:2009gt,Vary:2025yqo}. We derive its Lorentz-covariant decomposition in terms of four generalized FFs, which are subsequently calculated from BLFQ light-front wave functions (LFWFs)~\cite{Xu:2023nqv}. The resulting generalized FFs allow us to obtain the transverse color force distributions in IPS through Fourier transformation. We further extract the twist-3 matrix element $e_2$ and explore its connection to the transverse color Lorentz force experienced by a polarized quark inside the nucleon.

{\it Transverse color Lorentz force}.---
We begin by considering the local quark operators
\[
\bar{\psi}\,\Gamma\, i\overleftrightarrow{D}^{\{\mu_1}
i\overleftrightarrow{D}^{\mu_2\}}\psi ,
\]
where $\psi$ denotes the quark field, $\Gamma$ represents a Dirac matrix structure, $\overleftrightarrow{D}=(\overrightarrow{D}-\overleftarrow{D})/2=(\overrightarrow{\partial}-\overleftarrow{\partial})/2-igA$, and $a^{\{\mu}b^{\nu\}}=(a^\mu b^\nu+a^\nu b^\mu)/2$ and $a^{[\mu}b^{\nu]}=(a^\mu b^\nu-a^\nu b^\mu)/2$ defines the symmetrization and anti-symmetrization respectively. In this work, we focus on the scalar case with $\Gamma=1$. By inserting the identity operator $(\gamma^+\gamma^-+\gamma^-\gamma^+)/2$ and applying the equation of motion (EOM),
\[
(i\gamma^\mu D_\mu-m)\psi=0,
\]
the scalar operator can be expressed in terms of a quark--gluon correlation operator as
\begin{equation}\label{ope}
    \bar{\psi}\, i\overleftrightarrow{D}^{+}
    i\overleftrightarrow{D}^{+}\psi
    =
    -\frac{g}{2}\sum_{i=1}^{2}
    \bar{\psi}\,\sigma^{+i}G^{+i}\psi
    +m\,\bar{\psi}\gamma^+ i\overleftrightarrow{D}^{+}\psi .
\end{equation}
Here, we have used the identities $i\gamma^+\gamma^\perp=-\sigma^{+\perp}$, $\sigma^{\mu\nu}=i[\gamma^\mu,\gamma^\nu]/2$, and $[D^\mu,D^\nu]=-igG^{\mu\nu}$, where $G^{\mu\nu}$ is the gluon field-strength tensor. The first term on the right-hand side of Eq.~\eqref{ope} corresponds to the genuine twist-3 quark--gluon correlation that gives rise to the transverse color Lorentz force.

We now turn to the scalar twist-3 parton distribution function (PDF) $e(x)$ \cite{Jaffe:1991ra,Burkardt:2001iy}, which is associated with the $\Gamma=1$ quark bilinear. It is defined through the light-cone correlation function
\begin{equation}
    \int\frac{d\lambda}{2\pi}e^{i\lambda x}
    \langle P|\bar{\psi}(0)\psi(\lambda n)|P\rangle
    =2M e(x),
\end{equation}
where $n$ is the light cone basis
vector, which in the target rest frame is $n=(1,-1;\vec{0}_\perp)$.
Integrating both sides over $x$ leads to the sum rule
\begin{equation}
    \int dx\, e(x)
    =\frac{1}{2M}\langle P|\bar{\psi}(0)\psi(0)|P\rangle .
\end{equation}
The operator product expansion (OPE) relates the Mellin moments of $e(x)$ to its genuine twist-3 component $\bar e(x)$ and the twist-2 unpolarized PDF $f_1(x)$ as \cite{Burkardt:2001iy}
\begin{equation}
    \int dx\, x^n e(x)
    =
    \int dx\, x^n \bar e(x)
    +\frac{m}{M}\int dx\, x^{n-1}f_1(x).
\end{equation}
For $n=2$, combining this relation with Eq.~\eqref{ope} gives
\begin{equation}
\begin{aligned}
    &\int dx\,x^2\bar e(x)=
    -\frac{g}{4M(P^+)^2}
    \sum_{i=1}^{2}
    \langle P|
    \bar{\psi}(0)\sigma^{+i}G^{+i}(0)\psi(0)
    |P\rangle .
\end{aligned}
\end{equation}
Defining $e_2=\int dx\,x^2\bar e(x),$
we recover the relation introduced in Ref.~\cite{Burkardt:2008ps}. The matrix element on the right-hand side can be interpreted in terms of the transverse color Lorentz force by expressing the gluon field-strength tensor in light-cone coordinates,
\begin{equation}
    G^{+j}\sim E^j-\epsilon_T^{ij}B^i
    =(\vec E+\vec v\times\vec B)^j .
\end{equation}
Moreover, the presence of the tensor structure $\sigma^{+j}$ indicates that the quark is transversely polarized \cite{Diehl:2005jf,Burkardt:2005hp}. Since $e_2$ is independent of the overall polarization of the nucleon, it provides a unique characterization of the average transverse color Lorentz force experienced by a transversely polarized quark inside an unpolarized nucleon \cite{Burkardt:2008ps,Abdallah:2016xfk}.

To explore the spatial structure of the transverse color force, we parameterize the relevant matrix elements in terms of Lorentz-invariant FFs. Their Fourier transforms yield IPS distributions, providing a spatial picture of the transverse color forces inside the nucleon~\cite{burkardt2002impact,Aslan:2019jis,crawford2025transverse,liu2025color,Zhang:2026cig}.

\begin{figure*}[htbp]
    \centering
    \subfloat[ ]{\includegraphics[scale=0.6]{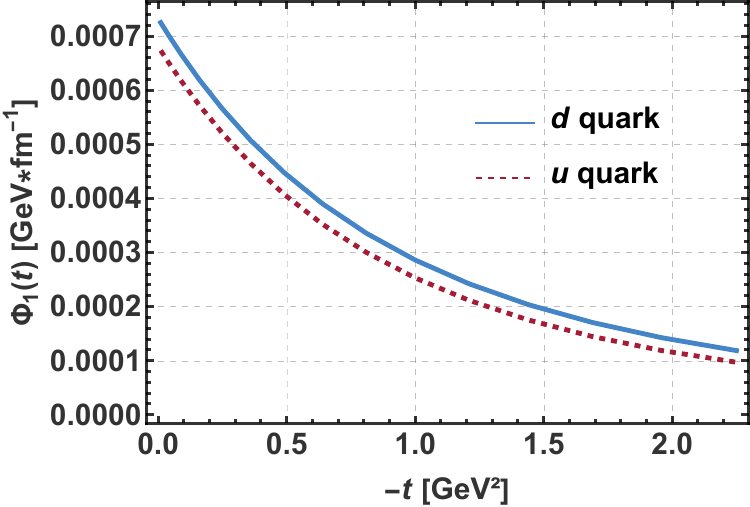}} \hspace{.3in}
    \subfloat[ ]{\includegraphics[scale=0.6]{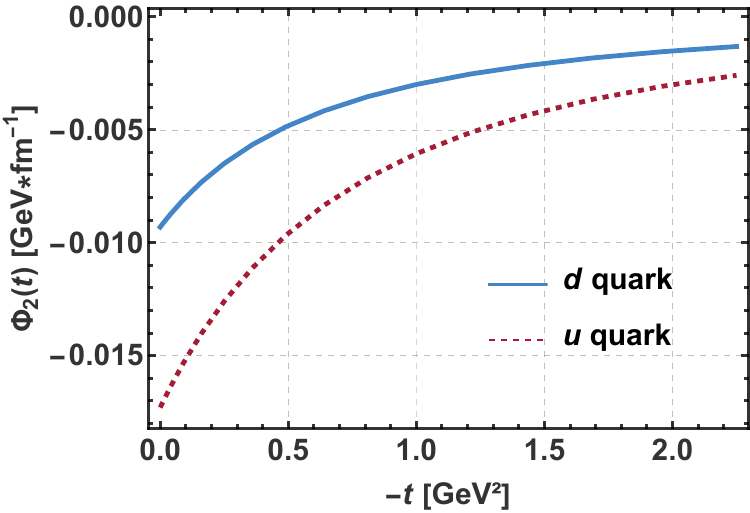}} \\
    \subfloat[ ]{\includegraphics[scale=0.6]{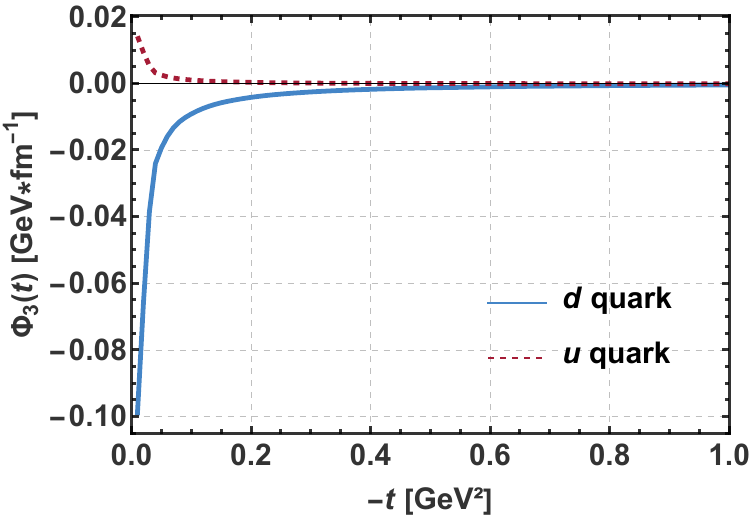}} \hspace{.3in}
    \subfloat[ ]{\includegraphics[scale=0.6]{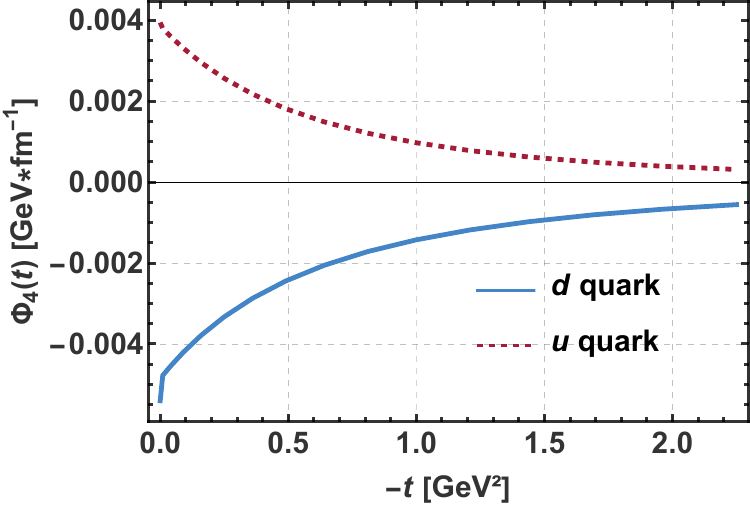}}
    \captionsetup{justification=raggedright}
    \caption{Four form factors $\Phi_1, \Phi_2, \Phi_3$ and $\Phi_4$ plotted as a function of $-t$ for quarks inside a proton in momentum space. The blue solid line stands for the $d$ quark, and the red dashed line for two $u$ quarks, both at the model initial scale.}
    \label{momspace}
\end{figure*}

\begin{figure*}[htbp]
    \centering
    \subfloat[ ]{\includegraphics[scale=0.3]{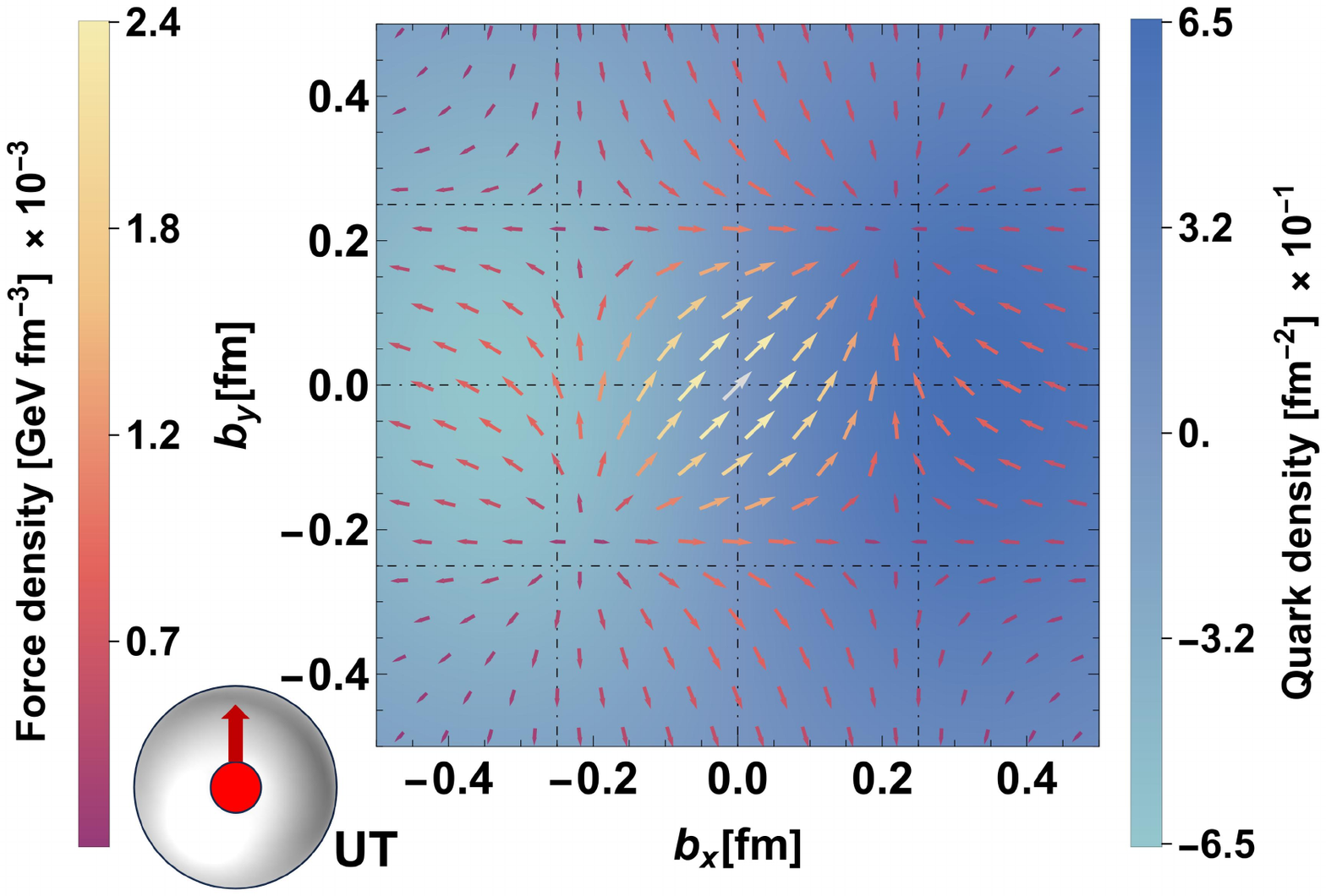}} \hspace{.0in}
    \subfloat[ ]{\includegraphics[scale=0.3]{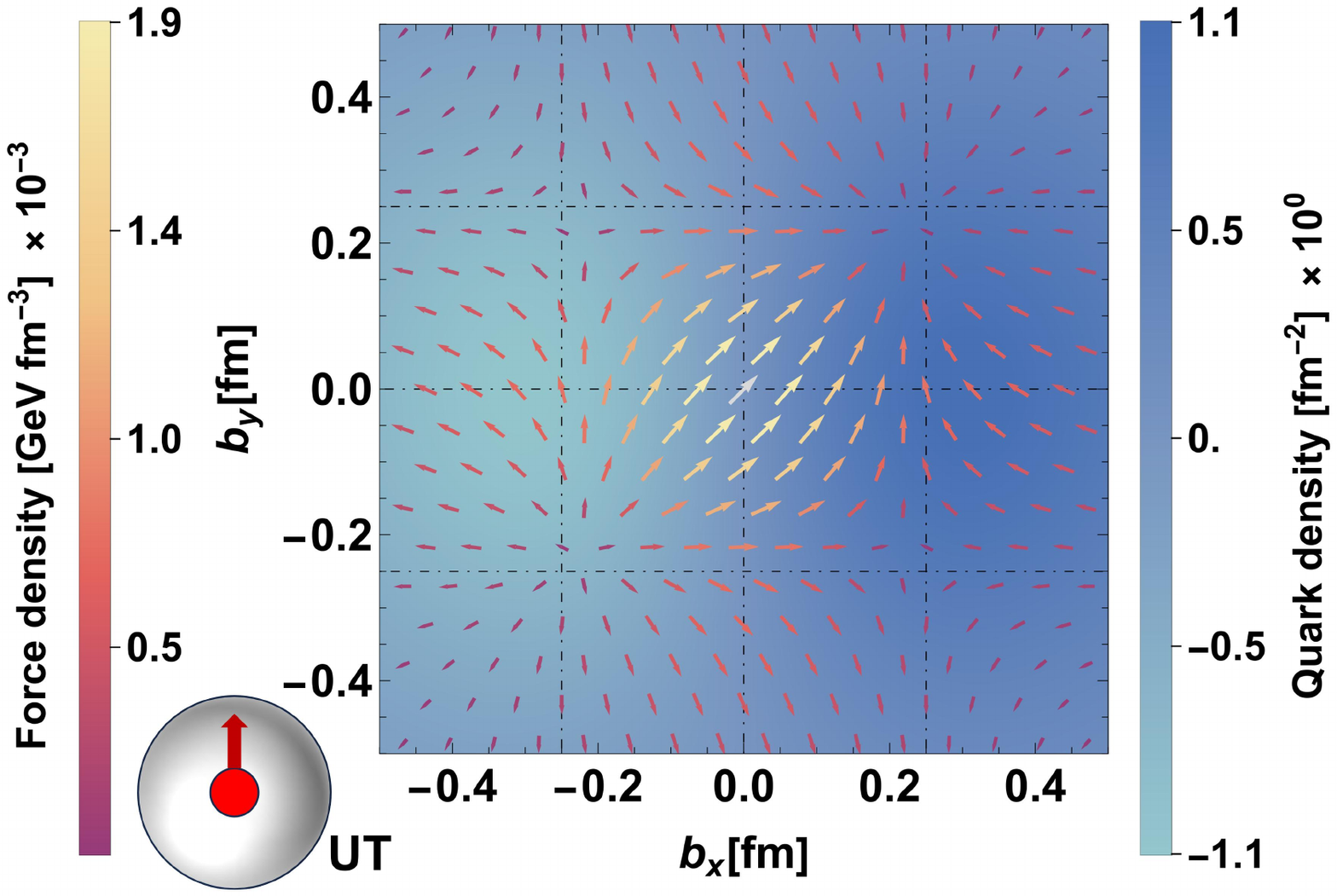}} \\
    \subfloat[ ]{\includegraphics[scale=0.3]{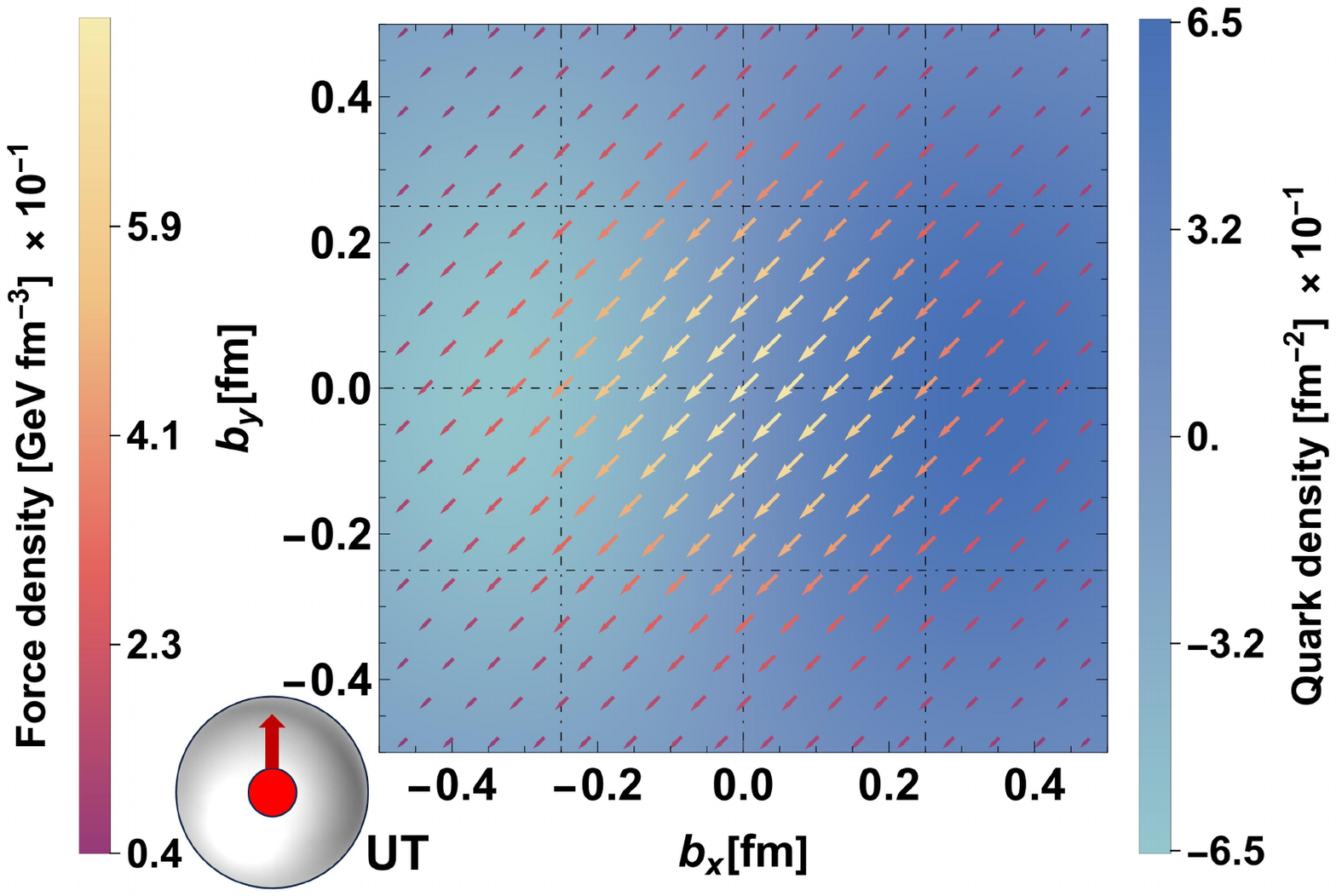}} \hspace{.0in}
    \subfloat[ ]{\includegraphics[scale=0.3]{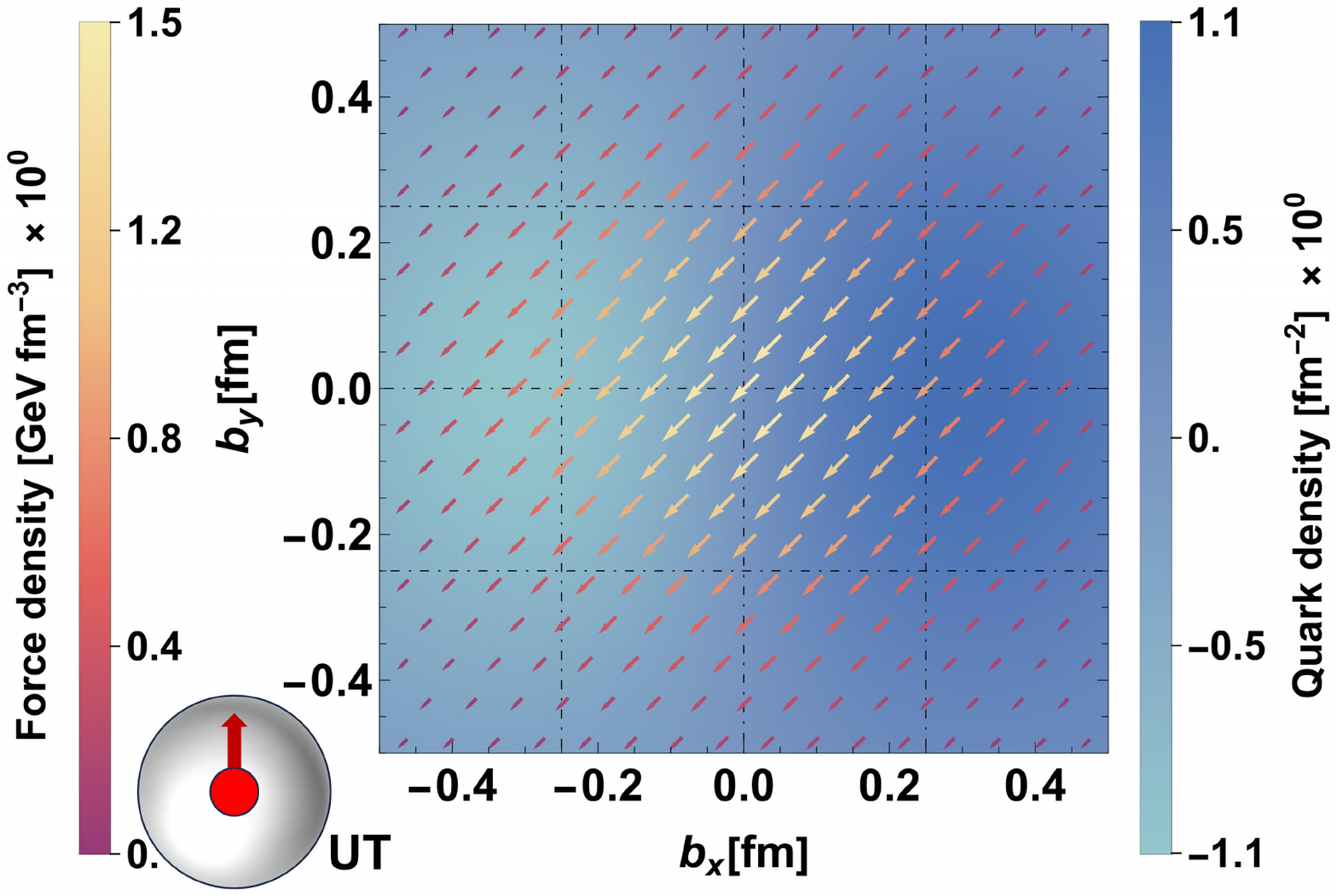}}
    \captionsetup{justification=raggedright}
    \caption{Distributions of the force components $\mathcal{F}_1$ and $\mathcal{F}_2$ in IPS, overlaid with the corresponding quark density distributions. The left and right panels show the $d$ and $u$ quark results, respectively. From top to bottom, the panels correspond to $\mathcal{F}_1$ and $\mathcal{F}_2$. The background is generated by a quark polarized along the $y$-axis.}
    \label{plotforce12}
\end{figure*}

\begin{figure*}[htbp]
    \centering
    \subfloat[ ]{\includegraphics[scale=0.3]{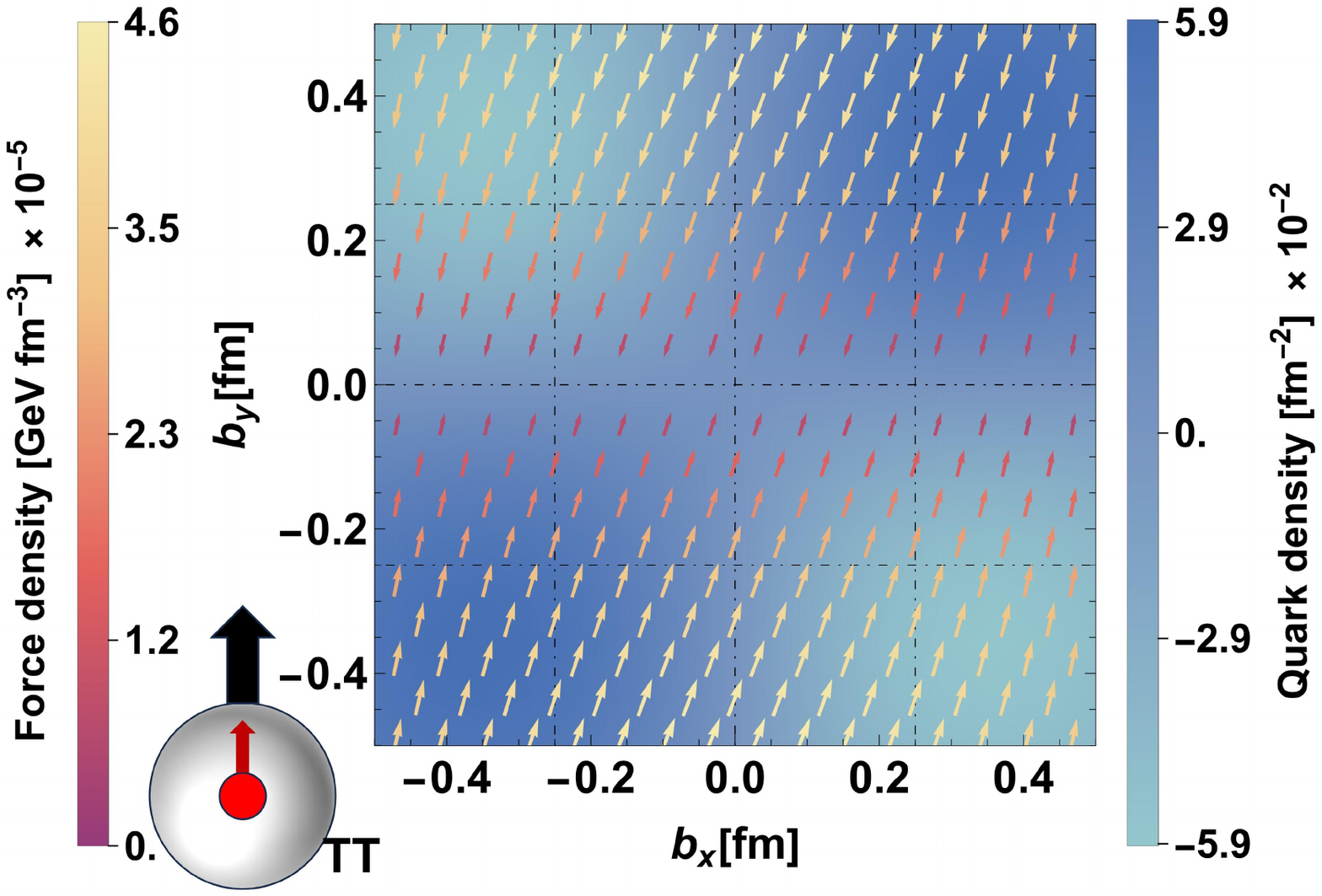}} \hspace{.0in}
    \subfloat[ ]{\includegraphics[scale=0.3]{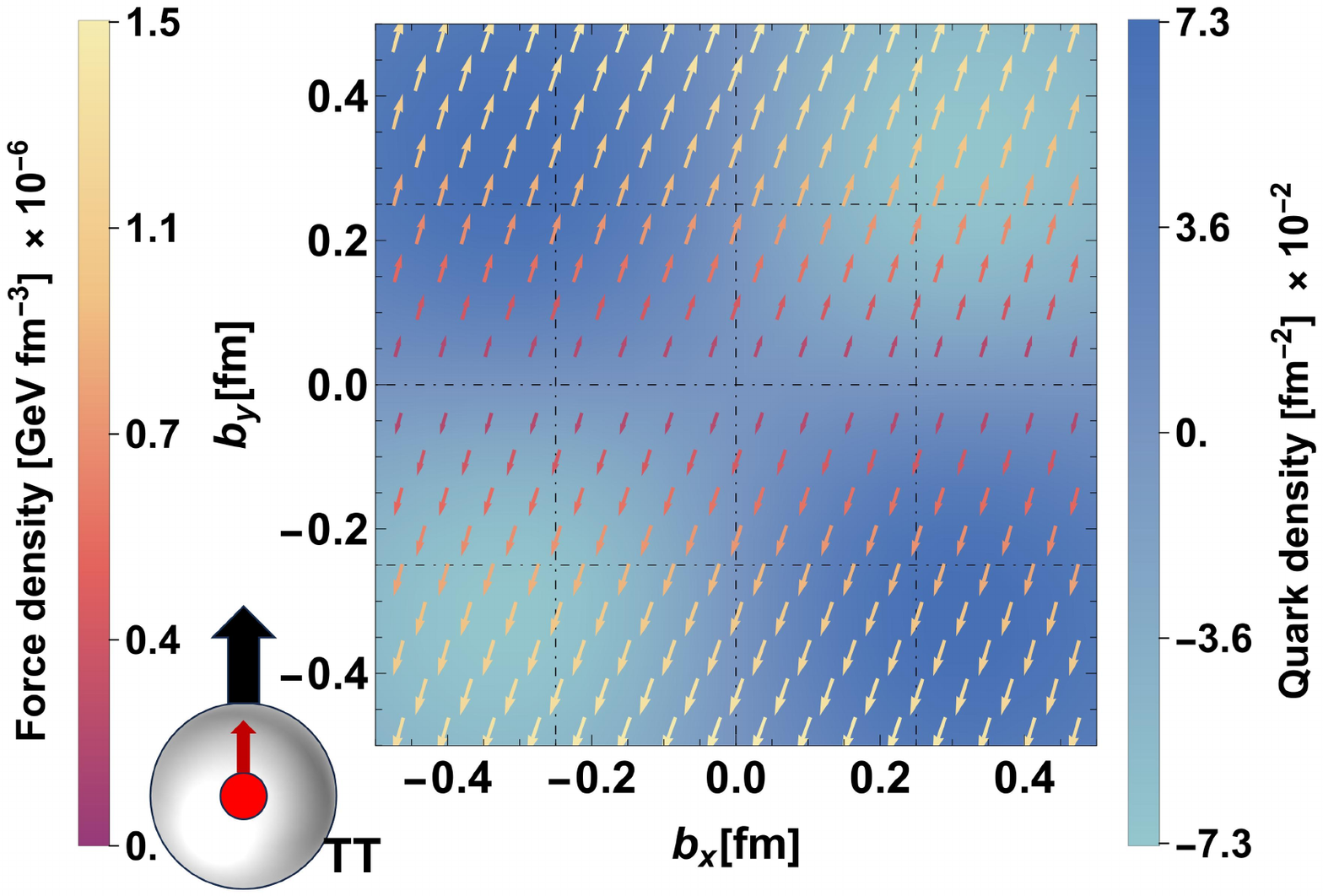}} \\
    \subfloat[ ]{\includegraphics[scale=0.3]{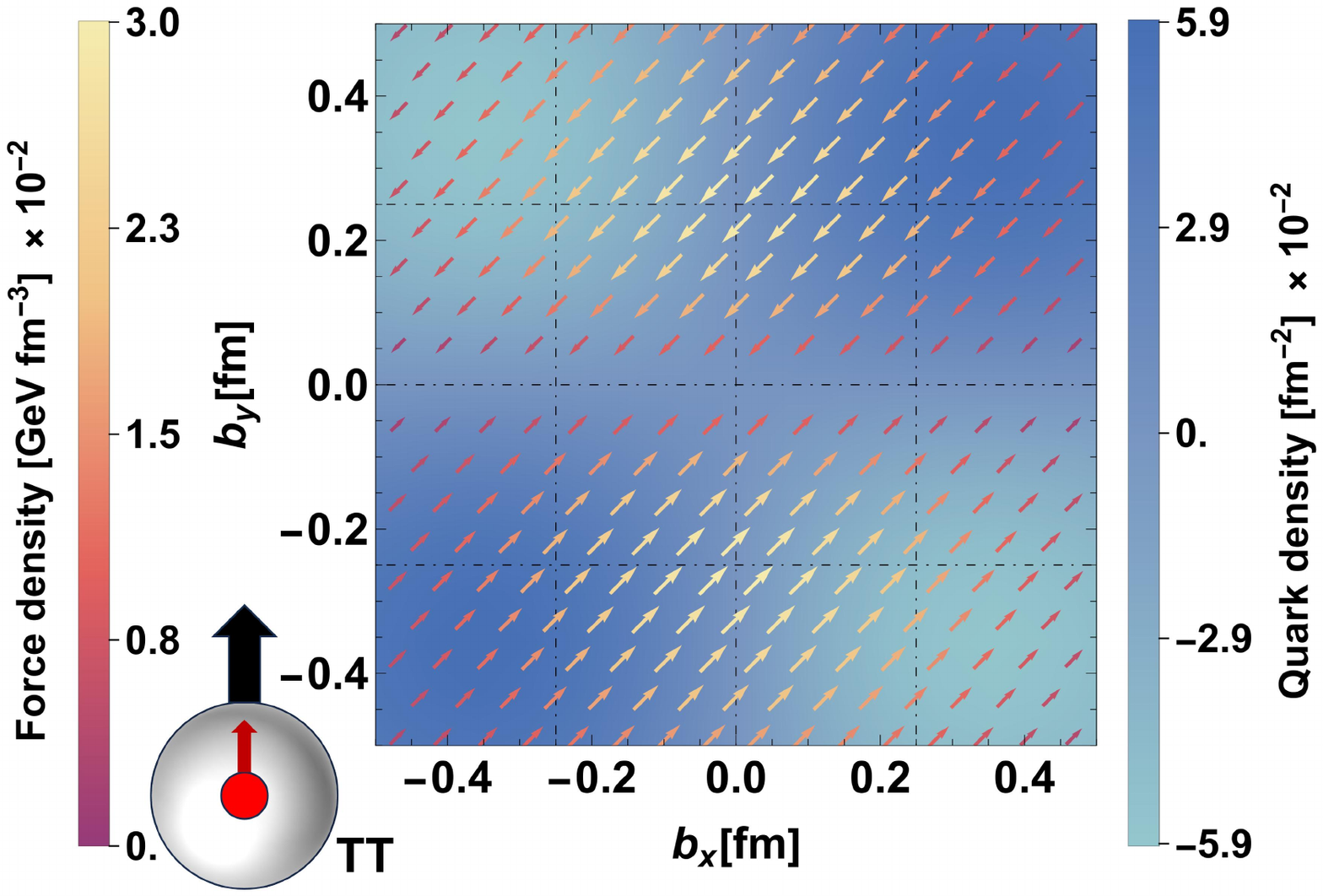}} \hspace{.0in}
    \subfloat[ ]{\includegraphics[scale=0.3]{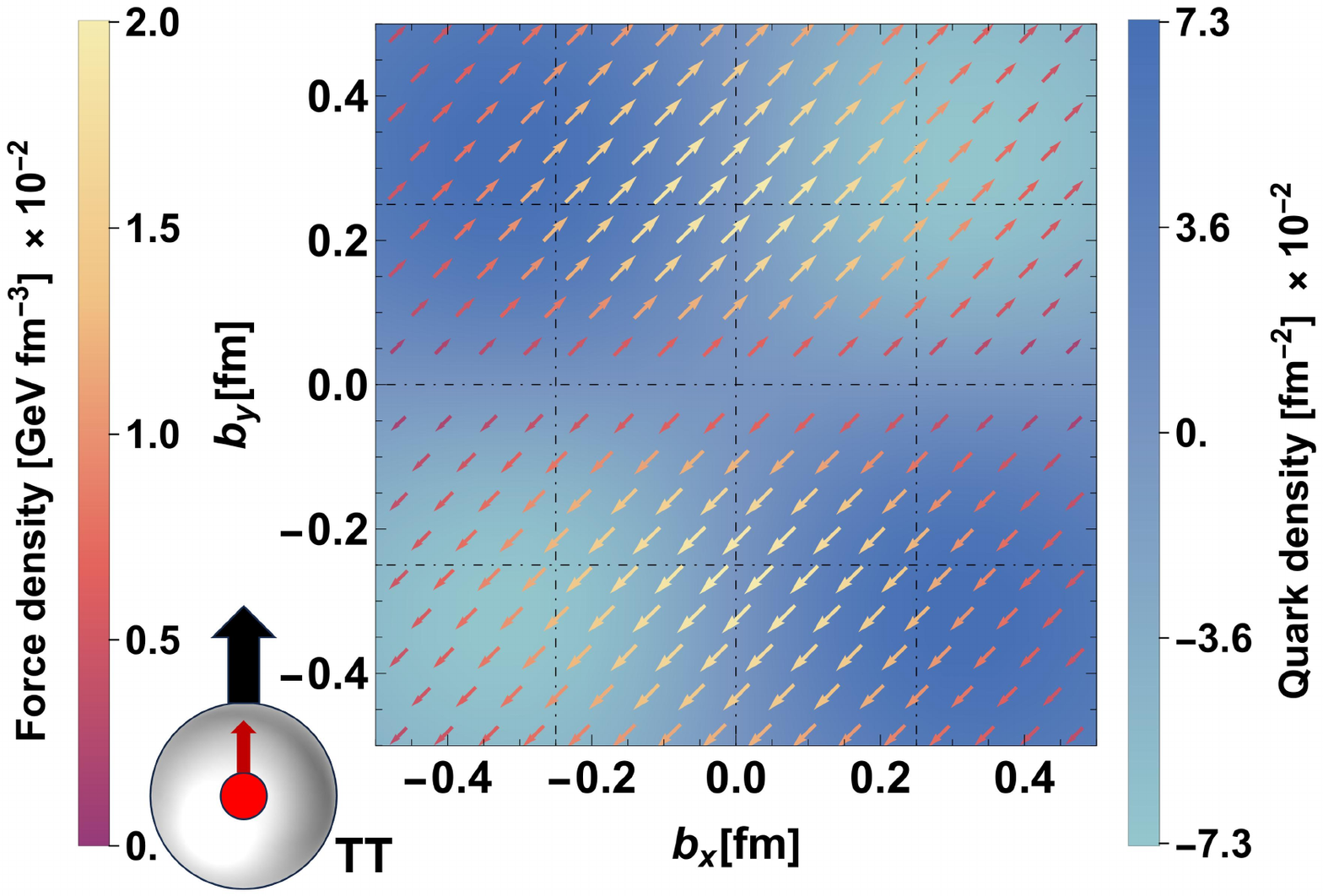}}
    \captionsetup{justification=raggedright}
    \caption{Distributions of the force components $\mathcal{F}_3$ and $\mathcal{F}_4$ in IPS, overlaid with the corresponding quark density distributions. The left and right panels show the $d$ and $u$ quark results, respectively. From top to bottom, the panels correspond to $\mathcal{F}_3$ and $\mathcal{F}_4$. The background is generated by a quark polarized along the $y$-axis and a proton polarized along the $x$-axis.}
    \label{plotforce34}
\end{figure*}
{\it Generalized form factors}.---We generalize the forward matrix element associated with $e_2$ to non-forward kinematics through the definition of an off-forward correlation function
\begin{equation}
    W^{\mu\nu, \rho\tau}_{\Lambda^\prime \Lambda} (p^\prime, p) = \langle p',\Lambda' | \bar{\psi}(0) \sigma^{\mu\nu} g G^{\rho\tau}(0) \psi(0) | p, \Lambda \rangle,
\end{equation}
and parameterize it as (see SUPPLEMENTAL MATERIAL for details)
    \begin{align}
    & W^{+j,+j}_{\Lambda^\prime \Lambda} (p^\prime, p) 
    = \bar{u}(p',\Lambda') \Big[\bar{P}^+\gamma^+\Big(\frac{\Delta^j\Delta^j}{M^3}\Phi_1(t)+\frac{g^{jj}}{M}\Phi_2(t)\Big) \nonumber \\&+\bar{P}^+i\sigma^{+\rho}\Delta_\rho\Big(\frac{\Delta^j\Delta^j}{M^4}\Phi_3(t)+\frac{g^{jj}}{M^2}\Phi_4(t)\Big) \Big]u(p,\Lambda),
\end{align}
where $p$ ($p'$) and $\Lambda$ ($\Lambda'$) denote the initial (final) proton momentum and helicity, respectively; $\bar{P}=(p+p')/2$, $\Delta=p'-p$, and $\Phi_i(t)$ ($i=1,\ldots,4$) represents the generalized FFs as a function of $t=-\Delta^2$.
The  color Lorentz force is defined as
\begin{equation} \label{ftdef}
    F^j_{\lambda^\prime \lambda} (t) = \frac{1}{2\sqrt{2} P^+} W^{+j, +j}_{\lambda' \lambda} (p', p).
\end{equation}
Note that this expression differs by a factor of two from the case of an unpolarized quark inside a transversely polarized proton \cite{Aslan:2019jis,Burkardt:2008ps,Zhang:2026cig}, reflecting the difference between a polarized quark state and an unpolarized quark ensemble. The corresponding transverse spatial distribution is obtained by performing the two-dimensional Fourier transform of Eq.~\eqref{ftdef}:
\begin{equation}
    \mathscr{F}^{j}_{\lambda^\prime \lambda} (\bm{b}_\perp) = \int \frac{d^2\Delta_\perp}{(2\pi)^2} e^{-i \bm{b}_\perp \cdot \bm{\Delta}_\perp} F^j_{\lambda^\prime \lambda} (\bm{\Delta_\perp}).
\end{equation}

We now present the LFWF overlap representations of the generalized FFs $\Phi_i(t)$:
\begin{align}
    \frac{(\Delta^j)^2}{M^2}\Phi_1(t)+\Phi_2(t)
    &= \int [dx]_4\, \frac{M[\sigma^{+j}]}{\sqrt{2}(P^+)^2}
    [\psi^{\uparrow\uparrow}_j], \label{phi1}\\
    \frac{(\Delta^j)^2}{M^2}\Phi_3(t)+\Phi_4(t)
    &= \int [dx]_4\, \frac{M^2[\sigma^{+j}]}{\sqrt{2}(P^+)^2\Delta^*}
    [\psi^{\uparrow\downarrow}_j], \label{phi2}
\end{align}
where
$
[\psi^{\Lambda'\Lambda}_j]
= \varepsilon_j \psi^{\Lambda'*}_{3}\psi^{\Lambda}_{4}
-\varepsilon_j^* \psi^{\Lambda'*}_{4}\psi^{\Lambda}_{3},
$
with $\varepsilon_j$ denoting the gluon polarization vector. The quantity
$[\sigma^{+j}]=\bar u(p',\Lambda')\sigma^{+j}u(p,\Lambda)$ contains the helicity structure of the struck quark. Here, $\psi_n^\Lambda$ represents the $n$-parton LFWF,
$\psi^\Lambda_{i=1,\cdots,n}(x_i,p_i,\lambda_i)$, for a proton with helicity $\Lambda$ and parton helicities $\lambda_i$. Specifically, $\psi_3^\Lambda$ and $\psi_4^\Lambda$ correspond to the $|qqq\rangle$ and $|qqqg\rangle$ Fock sectors, respectively. The helicity conservation of spectator partons,
$\delta^{\lambda'_2}_{\lambda_2}\delta^{\lambda'_3}_{\lambda_3}\delta^{\lambda'_4}_{\lambda_4}$, is implicitly understood. Finally, $\Delta=\Delta_1+i\Delta_2$ denotes the complex representation of the transverse momentum transfer.
We also introduce the following notations
\begin{align}
    [dx]_n &= \prod_{i=1}^n \frac{d x_{i} d^2 \vec{k}_{i}}{(16 \pi^3)^n} 16\pi^3 \delta \left(1-\sum x_i \right) \notag \\ 
    &\times \delta^2 \left(\sum \vec{k}_i \right) \delta (x-x_1),
\end{align}
where the subscript $1$ labels the struck quark.  The transverse index $j\in\{1,2\}$ labels the spatial component, and the choice of $j$ does not alter the $\Phi_i(t)$. 

By performing a two-dimensional Fourier transform, the corresponding IPS components of the transverse color Lorentz force distribution are derived as
\begin{align}
    \mathscr{F}^j_1 (\mathbf{b}_\perp) &= -\frac{P^+}{M^3} \left(\frac{\partial}{\partial b_j}\right)^2 \tilde{\Phi}_1 (b^2), \\
    \mathscr{F}^j_2 (\mathbf{b}_\perp) &= \frac{P^+}{M} \tilde{\Phi}_2 (b^2), \\
    \mathscr{F}^j_3 (\mathbf{b}_\perp) &= \frac{P^+}{M^4} \left[ \frac{\partial^3}{(\partial b_1)^2 \partial b_2}\delta^j_1 + \left(\frac{\partial}{\partial b_2}\right)^3 \delta^j_2 \right] \tilde{\Phi}_3 (b^2), \\
    \mathscr{F}^j_4 (\mathbf{b}_\perp) &= -\frac{P^+}{M^2} \frac{\partial}{\partial b_2} \tilde{\Phi}_4 (b^2),
\end{align}
where $\tilde{\Phi}_i$ represents the Fourier transform of the respective momentum-space FFs.

Physically, the FFs \(\Phi_1(t)\) and \(\Phi_2(t)\) emerge from the proton helicity-conserving (non-flip) transitions, as explicitly captured by the LFWF combination \([\psi^{\uparrow \uparrow}_j]\) in Eq.~\eqref{phi1}. Conversely, \(\Phi_3(t)\) and \(\Phi_4(t)\) require a proton helicity-flip transition, mediated by \([\psi^{\uparrow \downarrow}_j]\) in Eq.~\eqref{phi2}.

{\it Numerical results.}--- We present numerical results using proton LFWFs obtained by diagonalizing the proton light-front Hamiltonian within the BLFQ framework~\cite{Xu:2023nqv}, employing the $|qqq\rangle$ and $|qqqg\rangle$ Fock sectors. Figure~\ref{momspace} shows the momentum-space generalized FFs $\Phi_i(t)$ for up ($u$, red dashed) and down ($d$, blue solid) quarks. The squared momentum transfer range is $-t\in[0,2.25]$ GeV$^2$ for $\Phi_{1,2,4}$, while a narrower range $-t\in[0,1.0]$ GeV$^2$ is used for $\Phi_3$ to better display its detailed structure.

All four FFs exhibit an exponential decrease with increasing $-t$. For $\Phi_1$ (Fig.~\ref{momspace}a), both quark flavors are positive, with comparable magnitudes for the $u$ and $d$ quarks. In contrast, $\Phi_2$ (Fig.~\ref{momspace}b) is negative for both flavors, with the $u$-quark contribution being nearly twice as large as the $d$-quark one. Since $\Phi_1$ and $\Phi_2$ correspond to proton helicity-conserving transitions and are independent of the target polarization, they exhibit the same sign for both quark flavors.

In contrast, $\Phi_3$ (Fig.~\ref{momspace}c) exhibits a rapid decrease in the range $-t=0$--$0.1$ GeV$^2$, followed by a smooth approach toward zero. The $d$-quark contribution has a significantly larger magnitude than the $u$-quark contribution. For $\Phi_4$ (Fig.~\ref{momspace}d), the $u$- and $d$-quark contributions have comparable magnitudes. Since $\Phi_3$ and $\Phi_4$ arise from proton helicity-flip transitions and are sensitive to the transverse polarization of the target, they exhibit opposite signs for the $u$ and $d$ quark contributions.

In the forward limit, $\Delta\rightarrow0$, the FFs are related to the twist-3 matrix element $e_2^q$ through
\begin{equation}
    e_2^q = -\frac{2F (0)}{M^2} = -\frac{\Phi_2 (0)}{M^2}.
\end{equation}
Using the BLFQ LFWFs, we obtain $e_2^d=0.0102$ and $e_2^u=0.0189$ at the initial model scale $\mu_0^2=0.24\pm0.01~\mathrm{GeV}^2$. These values are comparable to the previously calculated twist-3 matrix elements $d_2^d=-0.00996$ and $d_2^u=0.0200$, which characterize the transverse color Lorentz force for unpolarized quarks inside a proton \cite{Zhang:2026cig}. Evolving the distributions to $\mu^2=5~\mathrm{GeV}^2$ using the ``snowflake'' package \cite{rodini2024numerical} gives $e_2^d=0.00509$ and $e_2^u=0.00945$. At the same scale, the corresponding $d_2$ values are $d_2^d=-0.00409$ and $d_2^u=0.00820$ \cite{Zhang:2026cig}. The evolved results exhibit a consistent hierarchy, $|e_2^q|>|d_2^q|$, for both $u$ and $d$ quarks, in agreement with previous studies \cite{Burkardt:2008ps,QCDSF:2006tkx,LHPC:2007blg,Burkardt:2007xm}.

At the initial BLFQ scale, $e_2^q$ and $d_2^q$ have comparable magnitudes, suggesting similar strengths of the underlying tensor and axial-vector quark--gluon correlations within the nonperturbative LFWFs. This similarity may originate from the common quark--gluon dynamics underlying both matrix elements. Since $e_2$ and $d_2$ are associated with chiral-odd and chiral-even quark--gluon operators, respectively, and follow different QCD evolution patterns, their relative magnitudes are not expected to remain unchanged under scale evolution. The evolved results are consistent with phenomenological expectations and lattice studies indicating stronger transverse-spin-dependent correlations in the chiral-odd sector \cite{Burkardt:2008ps,QCDSF:2006tkx,LHPC:2007blg,Burkardt:2007xm}. Our results thus demonstrate how QCD evolution enhances the distinction between these two twist-3 matrix elements.

In Figs.~\ref{plotforce12} and \ref{plotforce34}, we present components of the transverse force distribution in IPS, obtained by Fourier transforming the corresponding FFs. Following Refs.~\cite{crawford2025transverse,liu2025color,Zhang:2026cig}, these distributions are shown together with the transversely polarized quark-density profiles. The transversely polarized quark density used in this work is given by \cite{Diehl:2005jf}
\begin{align} \label{density}
    \sim& \ \frac{s^i}{M} \epsilon^{ij}_T \frac{\partial}{\partial b^j} \big(E_T+2\tilde{H}_T\big) + \frac{s^iS^j}{M^2}\frac{\partial^2}{\partial b^1 \partial b^2}\tilde{H}_T \notag \\
     &+\ s^iS^i\Big(2H_T-\frac{1}{M^2}\Big(\frac{\partial}{\partial b^i}\Big)^2 \tilde{H}_T\Big),
\end{align}
where $s^i$ and $S^i$ denote the transverse spin vectors of the quark and proton, respectively. The functions $H_T$, $E_T$, and $\tilde H_T$ are the twist-2 GPDs~\cite{Diehl:2001pm,Diehl:2005jf,Burkardt:2005hp}.

Figure~\ref{plotforce12} shows the IPS distributions of $\mathscr{F}_1$ and $\mathscr{F}_2$ (top and bottom rows, respectively), with the $d$- and $u$-quark results presented in the left and right panels. The $\mathscr{F}_1$ distribution exhibits a pronounced quadrupole-like pattern with reflection symmetry about the diagonal line $b_x=b_y$. The transverse plane is separated into regions with opposite force orientations by saddle-like transition regions. In contrast, $\mathscr{F}_2$ displays a simpler and more symmetric pattern, with force vectors predominantly aligned along the $b_x=b_y+c$ directions. Its magnitude is maximal near the center of the transverse plane and gradually decreases toward the periphery. This distribution can be interpreted as the IPS representation of the Boer--Mulders force, providing an intuitive spatial picture of the final-state interaction mechanism underlying the Boer--Mulders effect.

Figure~\ref{plotforce34} shows the IPS distributions of $\mathscr{F}_3$ and $\mathscr{F}_4$ (top and bottom rows), with $d$- and $u$-quark results in the left and right panels, respectively. The $\mathscr{F}_3$ distribution exhibits a shear-like flow with opposite force orientations in the upper and lower half-planes. This angular structure reflects the third-order transverse derivatives in $\mathscr{F}_3$, generating an octupole-like component in impact-parameter space. In contrast, $\mathscr{F}_4$ shows a dipole-like pattern associated with its first-order derivative structure. The force vectors are mainly oriented at approximately $45^\circ$ to the transverse axes, with their largest magnitudes near $(b_x,b_y)\approx(0,\pm0.25~\mathrm{fm})$. The $u$- and $d$-quark distributions exhibit similar spatial patterns but opposite force orientations and overall signs.

The background color maps show the transverse quark density, in units of $\mathrm{fm}^{-2}$, obtained from Eq.~(\ref{density}). For the unpolarized proton in Fig.~\ref{plotforce12}, the density is generated by the distortion term $\frac{s^i}{M}\epsilon_T^{ij}\frac{\partial}{\partial b^j}(E_T+2\tilde H_T)$, describing the transverse distortion of polarized quarks. For the transversely polarized proton in Fig.~\ref{plotforce34}, the density is determined by $\frac{s^iS^j}{M^2}\frac{\partial^2}{\partial b^1\partial b^2}\tilde H_T$, characterizing the spin-dependent transverse deformation from quark--proton spin correlations.

This connection is particularly relevant to transverse single-spin asymmetries. For a quark polarized along $+\hat{x}$ inside an unpolarized nucleon, the associated transverse-momentum correlation is described by the Boer--Mulders function $h_1^\perp(x,\mathbf{k}_\perp^2)$. Integrating over $x$ and taking the appropriate transverse-momentum moment relates the average transverse momentum along $+\hat{y}$ to the quark--gluon correlation as \cite{Burkardt:2008ps}
\begin{align}
    \langle k^y \rangle =& \int dx\int d^2\mathbf{k}_\perp \mathbf{k}_\perp^2 h_1^\perp(x,\mathbf{k}_\perp^2) \notag \\
    =& \frac{1}{2P^+} \int_0^\infty dx^- g\langle p|\bar{\psi}(0) \sigma^{+y} G^{+y}(x^-) \psi(0)|p\rangle.
\end{align}
The integration over $x^-$ follows the approximately light-like trajectory of the struck quark as it propagates away from the target, giving the integrand a natural interpretation as the transverse color Lorentz force along its trajectory. At $x^-=0$, it represents the force acting immediately after the quark absorbs the virtual photon. Thus, the local quark--gluon correlation entering the $x^2$ moment $e_2$ characterizes the instantaneous transverse force on a transversely polarized quark in an unpolarized nucleon. In this sense, $e_2$ is complementary to the Boer--Mulders function $h_1^\perp$, with the former describing the instantaneous force and the latter the accumulated transverse momentum from final-state interactions. Our calculation gives $\langle k^y\rangle_d=-0.0133$ and $\langle k^y\rangle_u=-0.0092$.

{\it Summary and conclusions.}--- We report the transverse color Lorentz force distributions experienced by a transversely polarized quark inside the nucleon, generated by the twist-3 quark--gluon correlation operator $\bar q \sigma^{+i}G^{+i}q$. Starting from the off-forward nucleon matrix element, we derive its complete Lorentz-covariant decomposition in terms of four generalized form factors, $\Phi_i(t)$ ($i=1,\ldots,4$), which are responsible for the spatial structure of the underlying quark--gluon correlations.

Using proton LFWFs obtained within the Basis Light-Front Quantization framework, we calculate these form factors and obtain their impact-parameter-space representations through Fourier transformation. This provides a spatial visualization of the transverse color-force distributions inside the proton. The resulting force components exhibit distinct geometric structures: $\mathscr{F}_4$ displays a dipole-like pattern, $\mathscr{F}_1$ exhibits a quadrupole-like structure, and $\mathscr{F}_3$ contains a dominant octupole-like component. In particular, $\mathscr{F}_2$, whose forward limit is directly related to the twist-3 matrix element $e_2$, provides an impact-parameter-space representation of the Boer--Mulders force and offers a spatial perspective on the final-state interaction mechanism associated with the Boer--Mulders effect. This matrix element is also connected to transverse single-spin asymmetries in SIDIS processes~\cite{Burkardt:2008ps}.

From the forward limits of the form factors, we extract the twist-3 matrix elements $e_2^q$ for both $u$ and $d$ quarks. At the initial scale, $e_2^q$ has a magnitude comparable to the corresponding $d_2^q$ matrix element, suggesting similar strengths of the underlying tensor and axial-vector quark--gluon correlations. After QCD evolution to $\mu^2=5~\mathrm{GeV}^2$, a clear hierarchy, $|e_2^q|>|d_2^q|$, emerges for both quark flavors. This behavior is consistent with phenomenological expectations and lattice-QCD studies indicating enhanced transverse-spin-dependent correlations in the chiral-odd sector. Overall, this work establishes a connection between twist-3 quark--gluon correlations, generalized FFs and transverse color-force distributions in impact-parameter-space, providing a new perspective on the three-dimensional structure of quark--gluon dynamics in the nucleon.\\

\noindent
{\bf Acknowledgments:} 
X. Zhao is supported by National Natural Science Foundation of China, Grant No. 12375143, by Key Research Program of Frontier Sciences, Chinese Academy of Sciences, Grant No.~ZDBS-LY-7020, by the Foundation for Key Talents of Gansu Province, by the Central Funds Guiding the Local Science and Technology Development of Gansu Province, Grant No.~22ZY1QA006, by international partnership program of the Chinese Academy of Sciences, Grant No.~016GJHZ2022103FN, and by the Strategic Priority Research Program of the Chinese Academy of Sciences, Grant No.~XDB34000000.
J. P. Vary is supported by the Department of Energy under Grant No.~DE-SC0023692.  A portion of the computational resources were also provided by Taiyuan Advanced Computing Center.

\bibliography{main}

@article{Xu:2023nqv,
    author = "Xu, Siqi and Mondal, Chandan and Zhao, Xingbo and Li, Yang and Vary, James P.",
    collaboration = "BLFQ",
    title = "{Quark and gluon spin and orbital angular momentum in the proton}",
    doi = "10.1103/PhysRevD.108.094002",
    journal = "Phys. Rev. D",
    volume = "108",
    number = "9",
    pages = "094002",
    year = "2023"
}

@article{Pacetti:2014jai,
    author = "Pacetti, S. and Baldini Ferroli, R. and Tomasi-Gustafsson, E.",
    title = "{Proton electromagnetic form factors: Basic notions, present achievements and future perspectives}",
    doi = "10.1016/j.physrep.2014.09.005",
    journal = "Phys. Rept.",
    volume = "550-551",
    pages = "1--103",
    year = "2015"
}

@article{Vary:2025yqo,
    author = "Vary, James P. and Mondal, Chandan and Xu, Siqi and Zhao, Xingbo and Li, Yang",
    collaboration = "BLFQ",
    title = "{Nucleon structure from basis light-front quantization: status and prospects}",
    eprint = "2512.08283",
    archivePrefix = "arXiv",
    primaryClass = "hep-ph",
    doi = "10.1140/epjs/s11734-025-02084-y",
    journal = "Eur. Phys. J. ST",
    volume = "235",
    number = "5",
    pages = "1557--1598",
    year = "2026"
}

@article{Burkert:2023wzr,
    author = "Burkert, V. D. and Elouadrhiri, L. and Girod, F. X. and Lorc{\'e}, C. and Schweitzer, P. and Shanahan, P. E.",
    title = "{Colloquium: Gravitational form factors of the proton}",
    eprint = "2303.08347",
    archivePrefix = "arXiv",
    primaryClass = "hep-ph",
    reportNumber = "JLAB-PHY-23-3774",
    doi = "10.1103/RevModPhys.95.041002",
    journal = "Rev. Mod. Phys.",
    volume = "95",
    number = "4",
    pages = "041002",
    year = "2023"
}

@article{Diehl:2023nmm,
    author = "Diehl, Stefan",
    title = "{Experimental exploration of the 3D nucleon structure}",
    doi = "10.1016/j.ppnp.2023.104069",
    journal = "Prog. Part. Nucl. Phys.",
    volume = "133",
    pages = "104069",
    year = "2023"
}

@article{Vary:2009gt,
  author        = {Vary, J. P. and Honkanen, H. and Li, Jun and Maris, P. and Brodsky, S. J. and Harindranath, A. and de Teramond, G. F. and Sternberg, P. and Ng, E. G. and Yang, C.},
  title         = {{Hamiltonian light-front field theory in a basis function approach}},
  eprint        = {0905.1411},
  archiveprefix = {arXiv},
  primaryclass  = {nucl-th},
  reportnumber  = {SLAC-PUB-13582},
  doi           = {10.1103/PhysRevC.81.035205},
  journal       = {Phys. Rev. C},
  volume        = {81},
  pages         = {035205},
  year          = {2010}
}

@article{crawford2025transverse,
  title={Transverse force distributions in the proton from lattice QCD},
  author={Crawford, JA and Can, KU and Horsley, R and Rakow, PE L and Schierholz, G and St{\"u}ben, H and Young, RD and Zanotti, JM and (QCDSF Collaboration)},
  journal={Physical Review Letters},
  volume={134},
  number={7},
  pages={071901},
  year={2025},
  publisher={APS}
}

@article{liu2025color,
  title={The color force acting on a quark in the pion and nucleon},
  author={Liu, Wei-Yang and Shuryak, Edward and Zahed, Ismail},
  journal={arXiv preprint arXiv:2511.09019},
  year={2025}
}

@article{Zhang:2026cig,
    author = "Zhang, Ziqi and Mondal, Chandan and Xu, Siqi and Zhao, Xingbo and Vary, James P.",
    collaboration = "BLFQ",
    title = "{Transverse force tomography inside a proton from basis light-front quantization}",
    eprint = "2603.25548",
    archivePrefix = "arXiv",
    primaryClass = "hep-ph",
    doi = "10.1103/sk7p-w8mg",
    journal = "Phys. Rev. D",
    volume = "114",
    number = "5",
    pages = "054011",
    year = "2026"
}

@article{burkardt2002impact,
  title={Impact parameter dependent parton distributions and transverse single spin asymmetries},
  author={Burkardt, Matthias},
  journal={Physical Review D},
  volume={66},
  number={11},
  pages={114005},
  year={2002},
  publisher={APS}
}

@article{rodini2024numerical,
  title={Numerical implementation of evolution equations for twist-3 collinear PDFs},
  author={Rodini, Simone and Rossi, Lorenzo and Vladimirov, Alexey},
  journal={The European Physical Journal C},
  volume={84},
  number={7},
  pages={732},
  year={2024},
  publisher={Springer}
}

@article{Aslan:2019jis,
    author = "Aslan, Fatma P. and Burkardt, Matthias and Schlegel, Marc",
    title = "{Transverse Force Tomography}",
    eprint = "1904.03494",
    archivePrefix = "arXiv",
    primaryClass = "hep-ph",
    doi = "10.1103/PhysRevD.100.096021",
    journal = "Phys. Rev. D",
    volume = "100",
    number = "9",
    pages = "096021",
    year = "2019"
}

@article{Burkardt:2008ps,
    author = "Burkardt, Matthias",
    title = "{Transverse force on quarks in deep-inelastic scattering}",
    eprint = "0810.3589",
    archivePrefix = "arXiv",
    primaryClass = "hep-ph",
    reportNumber = "JLAB-THY-08-912",
    doi = "10.1103/PhysRevD.88.114502",
    journal = "Phys. Rev. D",
    volume = "88",
    pages = "114502",
    year = "2013"
}

@article{Abdallah:2016xfk,
    author = "Abdallah, Manal and Burkardt, Matthias",
    title = "{Transverse Force on Transversely Polarized Quarks in Longitudinally Polarized Nucleons}",
    eprint = "1610.01166",
    archivePrefix = "arXiv",
    primaryClass = "hep-ph",
    doi = "10.1103/PhysRevD.94.094040",
    journal = "Phys. Rev. D",
    volume = "94",
    number = "9",
    pages = "094040",
    year = "2016"
}

@article{Jaffe:1991ra,
    author = "Jaffe, R. L. and Ji, Xiang-Dong",
    title = "{Chiral odd parton distributions and Drell-Yan processes}",
    reportNumber = "MIT-CTP-2005",
    doi = "10.1016/0550-3213(92)90110-W",
    journal = "Nucl. Phys. B",
    volume = "375",
    pages = "527--560",
    year = "1992"
}

@article{Burkardt:2001iy,
    author = "Burkardt, Matthias and Koike, Yuji",
    title = "{Violation of sum rules for twist three parton distributions in QCD}",
    eprint = "hep-ph/0111343",
    archivePrefix = "arXiv",
    doi = "10.1016/S0550-3213(02)00263-8",
    journal = "Nucl. Phys. B",
    volume = "632",
    pages = "311--329",
    year = "2002"
}

@article{Diehl:2005jf,
    author = "Diehl, M. and Hagler, Ph.",
    title = "{Spin densities in the transverse plane and generalized transversity distributions}",
    eprint = "hep-ph/0504175",
    archivePrefix = "arXiv",
    reportNumber = "DESY-05-061",
    doi = "10.1140/epjc/s2005-02342-6",
    journal = "Eur. Phys. J. C",
    volume = "44",
    pages = "87--101",
    year = "2005"
}

@article{Burkardt:2005hp,
    author = "Burkardt, Matthias",
    title = "{Transverse deformation of parton distributions and transversity decomposition of angular momentum}",
    eprint = "hep-ph/0505189",
    archivePrefix = "arXiv",
    doi = "10.1103/PhysRevD.72.094020",
    journal = "Phys. Rev. D",
    volume = "72",
    pages = "094020",
    year = "2005"
}

@article{Burkardt:2007xm,
    author = "Burkardt, Matthias and Hannafious, Brian",
    title = "{Are all Boer-Mulders functions alike?}",
    eprint = "0705.1573",
    archivePrefix = "arXiv",
    primaryClass = "hep-ph",
    doi = "10.1016/j.physletb.2007.09.064",
    journal = "Phys. Lett. B",
    volume = "658",
    pages = "130--137",
    year = "2008"
}

@article{Guo:2021aik,
    author = "Guo, Yuxun and Ji, Xiangdong and Shiells, Kyle",
    title = "{Novel twist-three transverse-spin sum rule for the proton and related generalized parton distributions}",
    eprint = "2101.05243",
    archivePrefix = "arXiv",
    primaryClass = "hep-ph",
    doi = "10.1016/j.nuclphysb.2021.115440",
    journal = "Nucl. Phys. B",
    volume = "969",
    pages = "115440",
    year = "2021"
}

@article{Diehl:2001pm,
    author = "Diehl, M.",
    title = "{Generalized parton distributions with helicity flip}",
    eprint = "hep-ph/0101335",
    archivePrefix = "arXiv",
    reportNumber = "DESY-01-009",
    doi = "10.1007/s100520100635",
    journal = "Eur. Phys. J. C",
    volume = "19",
    pages = "485--492",
    year = "2001"
}

@article{Hagler:2004yt,
    author = "Hagler, Ph.",
    title = "{Form-factor decomposition of generalized parton distributions at leading twist}",
    eprint = "hep-ph/0404138",
    archivePrefix = "arXiv",
    doi = "10.1016/j.physletb.2004.05.014",
    journal = "Phys. Lett. B",
    volume = "594",
    pages = "164--170",
    year = "2004"
}

@article{QCDSF:2006tkx,
    author = {G{\"o}ckeler, M. and H{\"a}gler, Ph. and Horsley, R. and Nakamura, Y. and Pleiter, D. and Rakow, P. E. L. and Sch{\"a}fer, A. and Schierholz, G. and St{\"u}ben, H. and Zanotti, J. M.},
    collaboration = "QCDSF, UKQCD",
    title = "{Transverse spin structure of the nucleon from lattice QCD simulations}",
    eprint = "hep-lat/0612032",
    archivePrefix = "arXiv",
    reportNumber = "DESY-06-245, EDINBURGH-2006-41, TUM-T39-06-16",
    doi = "10.1103/PhysRevLett.98.222001",
    journal = "Phys. Rev. Lett.",
    volume = "98",
    pages = "222001",
    year = "2007"
}

@article{LHPC:2007blg,
    author = "Hagler, Ph. and others",
    collaboration = "LHPC",
    title = "{Nucleon Generalized Parton Distributions from Full Lattice QCD}",
    eprint = "0705.4295",
    archivePrefix = "arXiv",
    primaryClass = "hep-lat",
    reportNumber = "DESY-07-077, JLAB-THY-07-651, TUM-T39-07-09",
    doi = "10.1103/PhysRevD.77.094502",
    journal = "Phys. Rev. D",
    volume = "77",
    pages = "094502",
    year = "2008"
}

\section*{SUPPLEMENTAL MATERIAL}\label{supp}
We begin with a general representation of this matrix structure, denoted by $\Gamma^{\mu\nu,\alpha\beta}$, where the Lorentz index pairs $(\mu,\nu)$ and $(\alpha,\beta)$ form antisymmetric pairs that represent $\sigma$ and $G$ separately.

Following the methodology adopted in Ref. \cite{Guo:2021aik}, we construct the building blocks from vector candidates. These are built from Dirac scalars, $V^\mu_S\in\{\bar{P}^\mu,\Delta^\mu,n^\mu\}$, where $n^\mu=\{0,1/\mathscr{P},0,0\}$ represents a light-cone vector pointing along the $-$ direction, as well as from Dirac tensors, $V^\mu_T\in\{\sigma^{\mu\nu}V_{S\nu}, V_S^\mu\sigma^{\alpha\beta}V_{S\alpha}V_{S\beta}\}$. For future convenience, we define the scalar-tensor operator as $S_T=\sigma^{\alpha\beta}V_{S\alpha}V_{S\beta}$. Utilizing this foundational basis along with the invariant tensors $\{g^{\mu\nu},\epsilon^{\mu\nu}_T,\sigma^{\mu\nu}\}$, we can now systematically construct $\Gamma^{\mu\nu,\alpha\beta}$ as
\begin{align}
    \Gamma^{\mu\nu,\alpha\beta} \sim \{&V_S^{[\mu}V_S^{\nu]}V_S^{[\alpha}V_S^{\beta]},V_S^{[\mu}V_S^{\nu]}V_S^{[\alpha}V_S^{\beta]}S_T, \notag \\ &V_T^{[\mu}V_S^{\nu]}V_S^{[\alpha}V_S^{\beta]},V_S^{[\mu}V_S^{\nu]}V_S^{[\alpha}V_T^{\beta]}, \notag \\ &\sigma^{\mu\nu}V_S^{[\alpha}V_S^{\beta]}, V_S^{[\mu}V_S^{\nu]}\sigma^{\alpha\beta}, V_S^{[\mu}g^{\nu][\alpha}V_S^{\beta]}, \\ &V_S^{[\mu}g^{\nu][\alpha}V_S^{\beta]}S_T, V_T^{[\mu}g^{\nu][\alpha}V_S^{\beta]}, V_S^{[\mu}g^{\nu][\alpha}V_T^{\beta]} \}. \notag
\end{align}

In order to further simplify this construction, we restrict our attention to the symmetric frame where $\bar{P}^\perp=0$, and take the zero-skewness limit, $\Delta^+=0$. This kinematic constraint reduces the parameterization to six remaining FFs
\begin{align}
    \Gamma^{+j,+j} =& \bar{P}^+\gamma^+\left(\frac{\Delta^j\Delta^j}{M^3}\Phi_1^\prime+\frac{g^{jj}}{M}\Phi_2^\prime\right) \notag \\
    &+ \bar{P}^+i\sigma^{+\rho}\Delta_\rho\left(\frac{\Delta^j\Delta^j}{M^4}\Phi_3^\prime+\frac{g^{jj}}{M^2}\Phi_4^\prime\right) \notag \\
    &+ \frac{(\bar{P}^+)^2\Delta^j}{M^4}i\sigma^{j\rho}\Delta_\rho \Phi_5^\prime + \frac{\bar{P}^+\Delta^j}{M^2}i\sigma^{+j} \Phi_6^\prime.
\end{align}
However, not all of these FFs are independent. Due to parity and time reversal symmetries, the pre-factor associated with $\Phi_5^\prime$ is strictly forbidden \cite{Diehl:2001pm,Hagler:2004yt}. Furthermore, invoking the transverse rotational symmetry between the $j=1$ and $j=2$ directions, the contribution from $\Phi_6^\prime$ can be fully absorbed into $\Phi_4^\prime$. Consequently, the final parameterization yields
\begin{align}
    \Gamma^{+j,+j} =& \bar{P}^+\gamma^+\left(\frac{\Delta^j\Delta^j}{M^3}\Phi_1+\frac{g^{jj}}{M}\Phi_2\right) \notag \\
    &+ \bar{P}^+i\sigma^{+\rho}\Delta_\rho\left(\frac{\Delta^j\Delta^j}{M^4}\Phi_3+\frac{g^{jj}}{M^2}\Phi_4\right).
\end{align}

\end{document}